\documentclass[12pt]{article}
\usepackage{amsmath,amsfonts,amssymb}  
\usepackage{graphicx}

 \ifx\pdfoutput\undefined
      \relax
 \else
      \usepackage{epstopdf}
 \fi

\makeatletter
\def\numberbysection{\@addtoreset{equation}{section}
    \def\theequation{\thesection.\arabic{equation}}}
\makeatother

\numberbysection

\newcommand{\be}{\begin{eqnarray}}
\newcommand{\ee}{\end{eqnarray}}
\newcommand{\non}{\nonumber}

\newcommand{\sh}{\mathop{\rm sh}\nolimits}
\newcommand{\ch}{\mathop{\rm ch}\nolimits}

\newcommand{\bff}{\ensuremath{\mathsf{b}}}

\newcommand{\Gf}{\ensuremath{\mathsf{G}}}
\newcommand{\Pf}{\ensuremath{\mathsf{P}}}
\newcommand{\yf}{\ensuremath{\mathsf{y}}}

\def\th{\theta}

\def\chi{\chi}

\def\La{\Lambda}

\newcommand{\beq}{\begin{equation}}
\newcommand{\eeq}{\end{equation}}
\newcommand{\bea}{\begin{eqnarray*}}
\newcommand{\eea}{\end{eqnarray*}}
\newcommand{\beqa}{\begin{eqnarray}}
\newcommand{\eeqa}{\end{eqnarray}}

\begin{document}

\begin{titlepage}
\strut\hfill
\vspace{.5in}
\begin{center}

\LARGE On the NLIE of (inhomogeneous) open spin-1 XXZ\\ 
\LARGE chain with general integrable boundary terms\\[1.0in]
\large Rajan Murgan\footnote{email: rmurgan@gustavus.edu}\\[0.8in]
\large Physics Department,\\ 
\large Gustavus Adolphus College,\\ 
\large 800 West College Avenue, St. Peter, MN 56082 USA\\
      
\end{center}

\vspace{.5in}

\begin{abstract}
Starting from the $T$-$Q$ equations of the open spin-$1$ XXZ quantum spin chain with general integrable boundary terms, for values of the boundary parameters
which satisfy a certain constraint, we derive a set of nonlinear integral equations (NLIEs) for the  inhomogeneous open spin-$1$ XXZ chain. By taking the continuum
limit of these NLIEs, and working in analogy with the open spin-$1$ XXZ chain with diagonal boundary terms, we compute the boundary and the Casimir energies
of the corresponding supersymmetric sine-Gordon (SSG) model. We also present an analytical result for the effective central charge in the ultraviolet (UV) limit.  
  
\end{abstract}

\end{titlepage}

\setcounter{footnote}{0}

\section{Introduction}\label{sec:intro}
Due to applications in statistical mechanics and condensed matter physics, one dimensional spin systems have attracted much interests. These models have been subjected
to intensive studies over the years. One such model in particular is the XXZ quantum spin chain with boundaries. For example, various aspects of the open spin-$1/2$ XXZ chain 
have been actively investigated \cite{Ga}-\cite{MurNep}. Among others, its finite size properties of ground and excited states have received much attention 
over the years. This lattice model has also been used to study the ground and excited states of the corresponding field theoretical model i.e., the sine-Gordon (SG) model with 
boundaries \cite{GZ}, let it be with Dirichlet or general integrable boundary conditions \cite{AN}, \cite{LMSS}-\cite{Murgan}. As for the periodic case \cite{DdV}-\cite{DdV3}, NLIEs 
have also been derived from the corresponding Bethe ansatz solutions of the open spin-$1/2$ XXZ chains with nondiagonal boundary terms \cite{nondiagonal}-\cite{Nep3},\cite{MurNep} 
and used in such investigations \cite{AN, BPT, ABR, ABNPT, Murgan}, in particular, in studies involving the crucial ultraviolet (UV) and infrared (IR) limits.

The spin-$1$ XXZ chain which is related to the supersymmetric sine-Gordon (SSG) model \cite{ssg1}-\cite{ANS} has also gone through similar studies over the years.  
The NLIEs have been obtained for cases with the periodic boundary conditions \cite{ArpadRavSuz} and Dirichlet boundary conditions \cite{ANS}. They have been used 
to compute quantities such as $S$ matrices (both bulk and boundary), bulk and boundary energies, central charges and conformal dimensions. However,  as pointed out in 
\cite{ANS}, since the ground state of the critical spin-$1$ XXZ chain is described by a sea of approximate ``two-strings'',  the usual method \cite{DdV}-\cite{DdV3} of deriving these 
NLIEs from Bethe ansatz equations and the counting functions does not seem to work in this case. Fortunately, NLIEs can be derived \cite{Su1, Su} rather by exploiting the analyticity of the 
transfer matrix eigenvalues described by the $T$-$Q$ equations of the model. Such a method have been used to derive the desired NLIEs describing SSG models with 
periodic boundary conditions \cite{ArpadRavSuz} and the Dirichlet boundary conditions \cite{ANS} and used to calculate finite-size properties of ground and 
excited states.

Our motivation for carrying out this work comes from these works, where spin-$1$ XXZ chains have been associated with SSG models. Moreover, solutions for the 
open spin-$1$ XXZ chain with general integrable boundary terms \cite{IOZ2} have been recently proposed \cite{FNR}. Hence, using this solution, one will be 
able to derive a set of NLIEs for the inhomogeneous open spin-$1$ XXZ chain which at the continuum limit should describe a SSG model. And since such NLIEs from open 
spin-$1$ XXZ chain with general integrable boundary terms have not been derived, we plan to address this issue in this paper. We shall only consider the ground state
in this paper.   

The outline of the article is as follows. In section 2,  the Hamiltonian and $T$-$Q$ equations for the open spin-$1$ XXZ spin chain with general integrable boundary 
terms are reviewed. This is followed by the derivation of a set of NLIEs for the inhomogeneous open spin-$1$ XXZ chain in section 3. We then give a brief review of the SSG 
model following \cite{NepSSG}. Proceeding by analogy with the SG model with general integrable boundary conditions \cite{AN, ABNPT} and the SSG model with Dirichlet boundary 
conditions \cite{ANS}, we consider the continuum limit of the NLIEs in section 4. In section 5, we compute the boundary and Casimir energies for the ground state. We 
analyze  the UV limit of the Casimir energy and give an analytical result for the effective central charge in section 6. Finally, we conclude the paper with a brief discussion of our results and some open problems in 
section 7.

\section{The open spin-$1$ XXZ chain}\label{sec:TQ}
\subsection{Hamiltonian of the open spin-$1$ XXZ chain}
In this section, we first begin by reviewing the Hamiltonian of the open spin-$1$ XXZ chain. We adopt the notations used in \cite{FNR}.

\be
{\cal H} = \sum_{n=1}^{N-1}H_{n,n+1} + H_{b} \,.
\label{Hamiltonianspin1}
\ee
where $H_{n,n+1}$ represents the bulk terms. Explicitly, these terms are given by \cite{ZF},
\be 
H_{n,n+1} &=&  \sigma_{n} - (\sigma_{n})^{2}
+ 2 \sh^2 \eta \left[ \sigma_{n}^{z} + (S^z_n)^2
+ (S^z_{n+1})^2 - (\sigma_{n}^{z})^2 \right] \non \\
&-& 4 \sh^2 (\frac{\eta}{2})  \left( \sigma_{n}^{\bot} \sigma_{n}^{z}
+ \sigma_{n}^{z} \sigma_{n}^{\bot} \right) \,, \label{bulkhamiltonianspin1}
\ee 
where
\be
\sigma_{n} = \vec S_n \cdot \vec S_{n+1} \,, \quad
\sigma_{n}^{\bot} = S^x_n S^x_{n+1} + S^y_n S^y_{n+1}  \,, \quad
\sigma_{n}^{z} = S^z_n S^z_{n+1} \,, 
\ee 
and $\vec S$ are the $su(2)$ spin-1 generators. $H_{b}$ represents the boundary terms which have the following form (see e.g., \cite{IOZ2})
\be 
H_{b} &=& a_{1} (S^{z}_{1})^{2}  + a_{2} S^{z}_{1} 
+  a_{3} (S^{+}_{1})^{2}  +  a_{4} (S^{-}_{1})^{2}  +
a_{5} S^{+}_{1}\, S^{z}_{1}  + a_{6}  S^{z}_{1}\, S^{-}_{1} \non \\
&+& a_{7}  S^{z}_{1}\, S^{+}_{1} + a_{8} S^{-}_{1}\, S^{z}_{1} 
+ (a_{j} \leftrightarrow b_{j} \mbox{ and } 1 \leftrightarrow N) \,,
\ee
where $S^{\pm} = S^{x} \pm i S^{y}$. The coefficients $\{ a_{i} \}$ 
of the boundary terms at site 1 are functions  
of the boundary parameters ($\alpha_{-}, \beta_{-},
\theta_{-}$) and the bulk anisotropy parameter $\eta$. They are given by,
\be
a_{1} &=& \frac{1}{4} a_{0} \left(\ch 2\alpha_{-} - \ch 
2\beta_{-}+\ch \eta \right) \sh 2\eta 
\sh \eta \,,\non \\
a_{2} &=& \frac{1}{4} a_{0} \sh 2\alpha_{-} \sh 2\beta_{-} \sh 2\eta \,, \non \\
a_{3} &=& -\frac{1}{8} a_{0} e^{2\theta_{-}} \sh 2\eta 
\sh \eta \,, \non \\
a_{4} &=& -\frac{1}{8} a_{0} e^{-2\theta_{-}} \sh 2\eta 
\sh \eta \,, \non \\
a_{5} &=&  a_{0} e^{\theta_{-}} \left(
\ch \beta_{-}\sh \alpha_{-} \ch {\eta\over 2} +
\ch \alpha_{-}\sh \beta_{-} \sh {\eta\over 2} \right)
\sh \eta \ch^{\frac{3}{2}}\eta \,, \non \\
a_{6} &=&  a_{0} e^{-\theta_{-}} \left(
\ch \beta_{-}\sh \alpha_{-} \ch {\eta\over 2} +
\ch \alpha_{-}\sh \beta_{-} \sh {\eta\over 2} \right)
\sh \eta \ch^{\frac{3}{2}}\eta \,, \non \\
a_{7} &=&  -a_{0} e^{\theta_{-}} \left(
\ch \beta_{-}\sh \alpha_{-} \ch {\eta\over 2} -
\ch \alpha_{-}\sh \beta_{-} \sh {\eta\over 2} \right)
\sh \eta \ch^{\frac{3}{2}}\eta \,, \non \\
a_{8} &=&  -a_{0} e^{-\theta_{-}} \left(
\ch \beta_{-}\sh \alpha_{-} \ch {\eta\over 2} -
\ch \alpha_{-}\sh \beta_{-} \sh {\eta\over 2} \right)
\sh \eta \ch^{\frac{3}{2}}\eta \,,
\ee
where 
\be
a_{0}= \left[
\sh(\alpha_{-}-{\eta\over 2})\sh(\alpha_{-}+{\eta\over 2})
\ch(\beta_{-}-{\eta\over 2})\ch(\beta_{-}+{\eta\over 2})\right]^{-1} 
\,.
\ee
Similarly, the coefficients $\{ b_{i} \}$ of the boundary terms at 
site $N$ which are functions of  
the boundary parameters ($\alpha_{+}, \beta_{+}, \theta_{+}$) and $\eta$, are given by the following correspondence,
\be
b_{i} = a_{i}\Big\vert_{\alpha_{-}\rightarrow \alpha_{+}, 
\beta_{-}\rightarrow -\beta_{+}, \theta_{-}\rightarrow \theta_{+}} \,.
\ee

\subsection{The $T$-$Q$ equations for the inhomogeneous open spin-$1$ XXZ chain}\label{sec:TQ}
Next, we review the $T$-$Q$ equations for the open spin-$1$ XXZ chain with general integrable 
boundary terms given in \cite{FNR}. However, we note that the solution holds only when the boundary parameters $\big(\alpha_{\pm}\,, \beta_{\pm}\,, \theta_{\pm}\big)$
obey the following constraint \cite{Nep}-\cite{Nep3},\cite{FNR},
\be
\alpha_{-} + \beta_{-} + \alpha_{+} + \beta_{+} = \pm (\theta_{-} - 
\theta_{+}) + \eta k \,,
\label{constraint}
\ee
where $k$ is an odd integer. A convenient redefinition of bulk and boundary parameters can be adopted \cite{AN}: 
\be
\eta = i \mu \,, \qquad 
\alpha_{\pm} = i \mu a_{\pm} \,, \qquad \beta_{\pm} = \mu b_{\pm} \,,
\qquad \theta_{\pm} = i \mu c_{\pm}
\label{newparams} \,,
\ee 
where $\mu\,, a_{\pm}\,, b_{\pm}\,, c_{\pm}$ are all real, 
with $0 < \mu < \frac{\pi}{2}$\footnote{Refer to section 4 of \cite{ANS} on IR limit for details on this bound.}.
With the above redefinitions, the constraint relation (\ref{constraint}) assumes the following pair of real contraints: 
\be
a_{-} + a_{+} &=& \pm |c_{-} - c_{+}| + k \,, \non \\
b_{-} + b_{+} &=& 0 \,.
\label{realconstraints}
\ee
In this paper, we shall consider only even $N$ case. Two relevant commuting transfer matrices for spin-$1$ XXZ chain are $T_1(u)$ with a spin-$\frac{1}{2}$ 
(two-dimensional) auxiliary space, and $T_2(u)$ with a spin-1 (three-dimensional) auxiliary space. The corresponding eigenvalues  
obey $T-Q$ equations found in \cite{FNR}: $\Lambda_1(u)$ which represents the eigenvalues of $T_{1}(u)$ can be written as 
(following \cite{ANS} and adopting the notations used there)
\beqa
\Lambda_1(u)&=&l_1(u)+l_2(u) \,, \non\\
l_1(u)&=&\sinh(2u+i\mu) \tilde{B}^{(+)}(u) \phi(u+i\mu) {Q(u-i\mu)\over Q(u)} 
\,, \non \\
l_2(u)&=&\sinh(2u-i\mu) \tilde{B}^{(-)}(u) \phi(u-i\mu) {Q(u+i\mu)\over Q(u)} 
\,, \label{T1} 
\eeqa
where 
\be
\phi(u)&=&\sinh^N(u-\La) \sinh^N(u+\La)\,, \non \\
\tilde{B}^{(\pm)}(u)&=&\sinh(u\pm \frac{i\mu A_{+}}{2}) \sinh(u\pm \frac{i\mu A_{-}}{2})\cosh(u\mp \frac{i\mu B_{+}}{2}) \cosh(u\mp \frac{i\mu B_{-}}{2}) 
\,,\non \\
Q(u)&=&\prod_{k=1}^{M}\sinh(u-\tilde{v}_k) \sinh(u+\tilde{v}_k) \,. 
\label{phiBpm}  
\ee
We have redefined the lattice boundary parameters as $A_{\pm} = 2a_{\pm} - 1\,, B_{\pm} = 2ib_{\pm} + 1$.In addition, $\La$ is the inhomogeneity parameter,
$N$ is the number of spins and $M = N-\frac{1}{2} + \frac{k}{2}$ represents the number of Bethe roots which are also the zeros $\tilde{v}_k$ of $Q(u)$. 
In this paper, we shall consider the case $k=1$. For this particular model of the XXZ chain, one generally needs to consider two groups of transfer matrix eigenvalues,
labelled as $\tilde{\Lambda}^{(\frac{1}{2},1)(\pm)}(u)$ in \cite{FNR}, to obtain all $3^{N}$ energy eigenvalues. Since we are interested only in the ground state, we restrict 
our analysis to only one of them that contains the ground state, namely $\tilde{\Lambda}^{(\frac{1}{2},1)(-)}(u)$. Readers are urged to refer to \cite{FNR} for details on this.
Next, using the fusion relation, one can write the eigenvalues of $T_{2}(u)$,  $\Lambda_2(u)$ as (see e.g. \cite{FNR})
\be
\Lambda_2(u) = \Lambda_1(u-{i\mu\over{2}})\, \Lambda_1(u+{i\mu\over{2}}) - f(u)
\label{T1T1}
\ee
where 
\be
f(u) &=& \phi(u+\frac{3i\mu}{2})\phi(u-\frac{3i\mu}{2})\sinh(2u-2i\mu)\sinh(2u+2i\mu)\tilde{B}^{(+)}(u+\frac{i\mu}{2})\tilde{B}^{(-)}(u-\frac{i\mu}{2})\non\\
&=&  l_1(u+\frac{i\mu}{2})l_2(u-\frac{i\mu}{2})
\label{f}
\ee
Using (\ref{f}), (\ref{T1T1}) can thus be written as 
\be
\Lambda_2(u) &=& l_2(u-\frac{i\mu}{2})l_2(u+\frac{i\mu}{2})+l_1(u-\frac{i\mu}{2})l_2(u+\frac{i\mu}{2})+l_1(u-\frac{i\mu}{2})l_1(u+\frac{i\mu}{2})\non \\
&=& \sinh (2u)\tilde{\Lambda}_2(u)
\label{T1T2}
\ee
where 
\be
\tilde{\Lambda}_2(u) &=& \sinh(2u-2i\mu) \tilde{B}^{(-)}(u-\frac{i\mu}{2}) \tilde{B}^{(-)}(u+\frac{i\mu}{2})\phi(u-\frac{3i\mu}{2}) \phi(u-\frac{i\mu}{2}){Q(u+\frac{3i\mu}{2})\over Q(u-\frac{i\mu}{2})} \non\\
&+&\sinh(2u) \tilde{B}^{(+)}(u-\frac{i\mu}{2}) \tilde{B}^{(-)}(u+\frac{i\mu}{2})\phi(u-\frac{i\mu}{2}) \phi(u+\frac{i\mu}{2}){Q(u+\frac{3i\mu}{2})\over Q(u-\frac{i\mu}{2})} {Q(u-\frac{3i\mu}{2})\over Q(u+\frac{i\mu}{2})}\non\\
&+&\sinh(2u+2i\mu) \tilde{B}^{(+)}(u+\frac{i\mu}{2}) \tilde{B}^{(+)}(u-\frac{i\mu}{2})\phi(u+\frac{3i\mu}{2}) \phi(u+\frac{i\mu}{2}){Q(u-\frac{3i\mu}{2})\over Q(u+\frac{i\mu}{2})} \non\\
&=& \tilde{\lambda}_{1}(u) + \tilde{\lambda}_{2}(u) + \tilde{\lambda}_{3}(u)
\label{Ttilde}
\ee
From (\ref{T1T2}) and (\ref{Ttilde}), we also have
\be
\tilde{\lambda}_{1}(u) = \frac{l_2(u-\frac{i\mu}{2})l_2(u+\frac{i\mu}{2})}{\sinh (2u)}\,,\quad \tilde{\lambda}_{2}(u) = \frac{l_1(u-\frac{i\mu}{2})l_2(u+\frac{i\mu}{2})}{\sinh (2u)}\,,\quad \tilde{\lambda}_{3}(u) = \frac{l_1(u-\frac{i\mu}{2})l_1(u+\frac{i\mu}{2})}{\sinh (2u)}
\non\\
\label{def}
\ee
One can now define the auxiliary functions $b(u)$ and $\bar{b}(u)$ by
\be
b(u) = {\tilde{\lambda}_{1}(u) + \tilde{\lambda}_{2}(u) \over \tilde{\lambda}_{3}(u)}\,, \qquad \bar{b}(u) = b(-u) = {\tilde{\lambda}_{3}(u) + \tilde{\lambda}_{2}(u) \over \tilde{\lambda}_{1}(u)}
\label{b}
\ee
We note that $\bar b(u)$ is the complex conjugate of $b(u)$ for real $u$. 
Using (\ref{T1}), (\ref{def}) and (\ref{b}), we obtain
\be
b(u) ={\Lambda_1(u-{i\mu\over{2}})\over{\sinh(2u+2i\mu)}}{\phi(u-{i\mu\over{2}})\over{
\phi(u+{i\mu\over{2}})\phi(u+{3i\mu\over{2}})}}
{\tilde{B}^{(-)}(u+{i\mu\over{2}})\over{\tilde{B}^{(+)}(u-{i\mu\over{2}})\tilde{B}^{(+)}(u+{i\mu\over{2}})}}
{Q(u+{3i\mu\over{2}})\over{Q(u-{3i\mu\over{2}})}}
\label{bb}
\ee
We also note that $\Lambda_{1}(u)$ does not have zeros near the real axis, except for a simple zero at the origin. Following \cite{ANS},
one can remove this root by defining,
\begin{equation}
\check \Lambda_{1}(u) = \frac{\Lambda_{1}(u)}{\kappa (u)}
\label{checklambda}
\end{equation}
where $\kappa (u)$ is any function whose only real root is a simple zero at the origin, that is $\kappa (0) = 0\,, \kappa '(0) \neq 0$. The prime
denotes differentiation with respect to $u$. In terms of  $\check \Lambda_{1}(u)$, (\ref{bb}) can be compactly written as,
\be
b(u)=C_b(u)\, \check \Lambda_1(u-{i\mu\over{2}})
{Q(u+{3i\mu\over{2}})\over{Q(u-{3i\mu\over{2}})}} \,,
\label{simplebandbarbT1Q}
\ee
where 
\be
C_b(u) &=&{\kappa (u-{i\mu\over{2}}) \phi(u-{i\mu\over{2}})\over{\sinh(2u+2i\mu)
\phi(u+{i\mu\over{2}})\phi(u+{3i\mu\over{2}})}}
{\tilde{B}^{(-)}(u+{i\mu\over{2}})\over{\tilde{B}^{(+)}(u-{i\mu\over{2}})\tilde{B}^{(+)}(u+{i\mu\over{2}})}} 
\label{Cb}
\ee
Next, defining 
\be
B(u) = 1+ b(u)\,, \qquad \bar{B}(u) = 1 + \bar{b}(u)\,, 
\ee
and using (\ref{Ttilde}) and (\ref{b}), we have the following expressions for $\tilde{\Lambda}_{2}(u)$,
\be
\tilde{\Lambda}_{2}(u) &=& \tilde{\lambda}_{3}(u)B(u) \non \\
&=& \tilde{\lambda}_{1}(u)\bar{B}(u) 
\label{newdeflambda}
\ee
As for $\Lambda_{1}(u)$ (see (\ref{checklambda})), $\tilde{\Lambda}_{2}(u)$ also has a root at the origin which can be removed by defining
\begin{equation}
\check \Lambda_{2}(u) = \frac{\tilde{\Lambda}_{2}(u)}{\kappa (u)}\,.
\label{Lamb2check}
\end{equation}
Using (\ref{T1}), (\ref{def}) and (\ref{newdeflambda}), (\ref{Lamb2check}) becomes
\be
\check \Lambda_{2}(u) &=& t_{+}(u){Q(u+{3i\mu\over{2}})\over{Q(u-{i\mu\over{2}})}}\bar{B}(u)\non \\
&=& t_{-}(u){Q(u-{3i\mu\over{2}})\over{Q(u+{i\mu\over{2}})}}B(u)
\label{lambdahat}
\ee
where
\be
t_{\pm}(u)={\sinh(2u\mp 2i\mu)\over \kappa(u)}
\tilde{B}^{(\mp)}(u-{i\mu\over{2}})\tilde{B}^{(\mp)}(u+{i\mu\over{2}})
\phi(u\mp{3i\mu\over{2}})\phi(u\mp{i\mu\over{2}}).
\label{tpm}
\ee
We next define the last two auxiliary functions $y(u)$ and $Y(u)$ as follows,
\be
y(u) = \frac{\sinh (2u)\tilde{\Lambda}_{2}(u)}{f(u)}\,, \qquad Y(u) = 1 + y(u),
\label{y}
\ee

\subsection{Bethe roots and parameter regions for ground state}\label{sec:Betheroot}

In this paper, we are primarily interested in studying the finite size properties of the ground state. Hence, the distribution of Bethe roots, $\big\{\tilde{v}_{k}\big\}$
for ground state of the open spin-$1$ XXZ chain described by the Hamiltonian (\ref{Hamiltonianspin1}), with the boundary parameters satisfying the constraint (\ref{constraint}), must first be given. 
The ground state is described by sea of approximate ``two-strings'', $\tilde{v}_{k} = x_{k} \pm iy_{k}$ (where $x_{k}$ and $y_{k}$ are real), the schematic depiction of which 
is given in figure 1 below. One can verify numerically that the imaginary parts $y_{k}$ satisfy $0 < y_{k}-\frac{\mu}{2} \ll 1$.

\setlength{\unitlength}{0.8cm}
\begin{picture}(5,3)(-3.5,-1.5)
\put(-2.5,0){\line(1,0){5}}
\put(0,-1.5){\line(0,1){3}}
\put(2.6,-0.1){$\small{\textrm{$u$}}$}
\put(0.0,1.30){$\tiny{\textrm{$i\mu/2$}}$}
\put(0.0,1.04){\circle*{0.1}}

\put(0.5,1.05){\circle*{0.1}}

\put(1.0,1.07){\circle*{0.1}}
\put(1.5,1.09){\circle*{0.1}}
\put(2.0,1.11){\circle*{0.1}}
\put(2.5,1.15){\circle*{0.1}}

\put(-0.5,1.05){\circle*{0.1}}

\put(-1.0,1.07){\circle*{0.1}}
\put(-1.5,1.09){\circle*{0.1}}
\put(-2.0,1.11){\circle*{0.1}}
\put(-2.5,1.15){\circle*{0.1}}

\put(0.0,-1.4){$\tiny{\textrm{$-i\mu/2$}}$}
\put(0.0,-1.04){\circle*{0.1}}

\put(0.5,-1.05){\circle*{0.1}}

\put(1.0,-1.07){\circle*{0.1}}
\put(1.5,-1.09){\circle*{0.1}}
\put(2.0,-1.11){\circle*{0.1}}
\put(2.5,-1.15){\circle*{0.1}}

\put(-0.5,-1.05){\circle*{0.1}}

\put(-1.0,-1.07){\circle*{0.1}}
\put(-1.5,-1.09){\circle*{0.1}}
\put(-2.0,-1.11){\circle*{0.1}}
\put(-2.5,-1.15){\circle*{0.1}}

\put(-1.8,-2.0){$\small{\textrm{Figure\, 1: Zeros of $Q(u)$}}$}
\end{picture}

\bigskip \noindent From numerical investigations, the regions in parameter space $A_{\pm}$ which yield the above form of Bethe roots for the ground state can be 
divided in the following way. \footnote{Such investigations have been given in \cite{AN} for the corresponding spin-$1/2$ XXZ chain. Readers are urged to refer to the reference for 
further details on such numerical investigations.}
\be
\begin{array}{r@{\ : \quad}l}
    I   & 1 < A_{\pm} < {2\pi\over \mu} \\
    II  & 1 < A_{+} < {2\pi\over \mu} \quad \& \quad 
    1-{2\pi\over \mu}< A_{-} < -1 \\
    III &  1-{2\pi\over \mu}< A_{\pm} < -1 \\
    IV  & 1-{2\pi\over \mu} < A_{+} < -1 \quad \& \quad 
    1< A_{-} < {2\pi\over \mu}
  \end{array}
  \label{regions}
\ee
In addition, due to (\ref{realconstraints}), the parameters $B_{\pm}$ satisfy 
\beq
B_{+} + B_{-} = 2
\label{consforB}
\eeq

\noindent In subsequent sections, following \cite{ANS}, we will work instead with a shifted function, defined by $q(u) = Q(u + \frac{i\pi}{2})$. This is due to the fact that
for $\mu\rightarrow 0$, $Q(u)$ can have zeros near the real axis (refer to figure 1). On the other hand, $q(u)$ does not suffer from such features. Refer to the 
figure 2 that gives schematic picture of zeros of $q(u)$.

\setlength{\unitlength}{0.8cm}
\begin{picture}(5,3)(-3.5,-1.3)
\put(-2.5,0){\line(1,0){5}}
\put(0,-1.5){\line(0,1){3}}
\put(2.6,-0.1){$\small{\textrm{$u$}}$}

\put(0.0,1.30){$\tiny{\textrm{$i(\pi-\mu)/2$}}$}
\put(0.0,1.13){\circle*{0.1}}

\put(0.5,1.11){\circle*{0.1}}

\put(1.0,1.09){\circle*{0.1}}
\put(1.5,1.07){\circle*{0.1}}
\put(2.0,1.05){\circle*{0.1}}
\put(2.5,1.03){\circle*{0.1}}

\put(-0.5,1.11){\circle*{0.1}}

\put(-1.0,1.09){\circle*{0.1}}
\put(-1.5,1.07){\circle*{0.1}}
\put(-2.0,1.05){\circle*{0.1}}
\put(-2.5,1.03){\circle*{0.1}}

\put(0.0,-1.4){$\tiny{\textrm{$-i(\pi-\mu)/2$}}$}
\put(0.0,-1.13){\circle*{0.1}}

\put(0.5,-1.11){\circle*{0.1}}

\put(1.0,-1.09){\circle*{0.1}}
\put(1.5,-1.07){\circle*{0.1}}
\put(2.0,-1.05){\circle*{0.1}}
\put(2.5,-1.03){\circle*{0.1}}

\put(-0.5,-1.11){\circle*{0.1}}

\put(-1.0,-1.09){\circle*{0.1}}
\put(-1.5,-1.07){\circle*{0.1}}
\put(-2.0,-1.05){\circle*{0.1}}
\put(-2.5,-1.03){\circle*{0.1}}

\put(-1.8,-2.5){$\small{\textrm{Figure\, 2: Zeros of $q(u)$}}$}
\end{picture}

\newpage \section{The NLIEs for the inhomogeneous open spin-$1$ XXZ chain}\label{sec:NLIE}

Utilizing the analyticity of $\ln \check{\Lambda}_2(u)$ near the real axis, we have the following from Cauchy's theorem,
\beq
0=\oint_C du\ [\ln\check{\Lambda}_2(u)]'' e^{iku}
\label{cauchy1}
\eeq
where the contour $C$ is chosen as in figure 3 below, $\epsilon$ is small and positive, such that the max $\{ y_{k}\} - \frac{\mu}{2}  < \epsilon$.

\setlength{\unitlength}{0.8cm}
\begin{picture}(6,4)(-9,-3)
\put(-2.5,0){\line(1,0){5}}
\put(0,-1.5){\line(0,1){3}}
\put(-1.0,0.3){\vector(-1,0){0.2}}
\put(-2.0,0.3){\line(1,0){4}}
\put(1.0,-0.3){\vector(1,0){0.2}}
\put(-2.0,-0.3){\line(1,0){4}}
\put(0.75,0.5){$\footnotesize{\textrm{$C_{1}$}}$}
\put(0.75,-0.8){$\footnotesize{\textrm{$C_{2}$}}$}
\put(2.0,-0.3){\line(0,1){0.6}}
\put(-2.0,-0.3){\line(0,1){0.6}}
\put(2.2,0.1){$\footnotesize{\textrm{$i\epsilon$}}$}
\put(2.2,-0.3){$\footnotesize{\textrm{$-i\epsilon$}}$}
\put(-2.0,-2.2){$\small{\textrm{Figure\, 3: Integration\, contour}}$}
\end{picture}
 
After using (\ref{lambdahat}), (\ref{cauchy1}) can be written as
\be
0 &=&\int_{C_1} du\ \left\{ \ln t_{-}(u) \right\}'' e^{iku}+
\int_{C_1} du\ \left\{ \ln \left[{q(u-\frac{3i\mu}{2} + \frac{i \pi}{2})\over 
q(u+\frac{i\mu}{2} - \frac{i \pi}{2})}\right] \right\}'' e^{iku} \non \\
&+&\int_{C_1} du\ \left\{ \ln B(u) \right\}'' e^{iku} + \int_{C_2} du\ \left\{ \ln t_{+}(u) \right\}'' e^{iku}\non \\
&+& \int_{C_2} du\ \left\{ \ln\left[{q(u+\frac{3i\mu}{2} - \frac{i \pi}{2})\over 
q(u-\frac{i\mu}{2} + \frac{i \pi}{2})}\right] 
\right\}'' e^{iku} + \int_{C_2} du\ \left\{ \ln \bar B(u)\right\}'' e^{iku} \,,
\label{cauchy1b}
\ee
As in \cite{ANS}, we define the following Fourier transforms along $C_{2}$ and $C_{1}$,
\be
\widehat{Lf''}(k)=\int_{C_2} du\ [\ln f(u)]'' e^{iku} \,, \qquad
\widehat{{\cal L}f''}(k)=\int_{C_1} du\ [\ln f(u)]'' e^{iku} \,,
\label{fouriertransfdef}
\ee
respectively. Exploiting the periodicity (to make the imaginary part of the argument negative),
\beq
q(u)=q(u-i\pi),\quad u\in C_1,\qquad{\rm and}\qquad
q(u+i\mu)=q(u+i\mu-i\pi),\quad u\in C_2 \,.
\label{qperiodicity}
\eeq
we obtain the following
\be
\int_{C_1} du\ \left\{ \ln \left[{q(u-\frac{3i\mu}{2} + \frac{i \pi}{2})\over 
q(u+\frac{i\mu}{2} - \frac{i \pi}{2})}\right] \right\}'' e^{iku}&=&
\widehat{Lq''}(k)\left(e^{(\frac{\mu}{2} - \frac{\pi}{2}) k}-e^{(\frac{\pi}{2} - \frac{3\mu}{2}) k}\right) \,, \non \\
\int_{C_2} du\ \left\{ \ln\left[{q(u+\frac{3i\mu}{2} - \frac{i \pi}{2})\over 
q(u-\frac{i\mu}{2} + \frac{i \pi}{2})}\right] 
\right\}'' e^{iku} &=&
\widehat{Lq''}(k)\left(e^{(\frac{3\mu}{2} - \frac{\pi}{2}) k}-e^{(\frac{\pi}{2} - \frac{\mu}{2}) k}\right) \,,
\label{c1c2}
\ee
Consequently, after using (\ref{fouriertransfdef}) for $t_{\pm}(u)$, $B(u)$ and $\bar B(u)$ together with (\ref{c1c2}), (\ref{cauchy1b}) becomes
\beq
\left[e^{\left({\pi\over{2}}-{3\mu\over{2}}\right)k}
-e^{\left({\mu\over{2}}-{\pi\over{2}}\right)k}
-e^{\left({3\mu\over{2}}-{\pi\over{2}}\right)k}
+e^{\left({\pi\over{2}}-{\mu\over{2}}\right)k}\right]\widehat{Lq''}(k)=
\widehat{L{\bar B}''}(k)+\widehat{{\cal L}B''}(k)+\widehat{Lt''_+}(k)+\widehat{{\cal L}t''_-}(k) \,,
\label{LqBB}
\eeq
Also from (\ref{simplebandbarbT1Q}), we have the following for the Fourier transform of $b(u)$, 
\beqa
\widehat{Lb''}(k)&=&e^{-\frac{\mu k}{2}}\widehat{L\check{\Lambda}''_1}(k)+\widehat{Lq''}(k)
\left[e^{\left({3\mu\over{2}}-{\pi\over{2}}\right)k}-
e^{\left({\pi\over{2}}-{3\mu\over{2}}\right)k}\right]+\widehat{LC''_b}(k) \,.
\label{LbLT1Lq}
\eeqa
In addition, from (\ref{T1T1}) and (\ref{checklambda}), we obtain\footnote{We remark that the term $2\pi k$ in (\ref{LT1Y})
is obtained by evaluating the integral $\oint_C du\ [\ln \kappa(u)]'' e^{iku}$.}
\beq
\left(e^{{\mu k\over{2}}}+e^{-{\mu k\over{2}}}\right)\widehat{L\check{\Lambda}''_1}(k)
=\widehat{LY''}(k) + \widehat{Lf''}(k) - \left(e^{\mu k\over 
2}+e^{-{\mu k\over 2}}\right) \widehat{L\kappa''}(k) + e^{\mu k\over 2} 2\pi k
\,.
\label{LT1Y}
\eeq
which can be used together with (\ref{LqBB}) to obtain the following NLIE in Fourier space for $b(u)$,
\beqa
\widehat{Lb''}(k)&=&-\widehat G(k)\left[\widehat{L{\bar B}''}(k)+\widehat{{\cal L}B''}(k)\right]+
\widehat G_2(k)\, \widehat{LY''}(k)+C(k) 
\label{Lbnlie}
\eeqa
where
\beqa
\widehat G(k)&=&
{\sinh\left((\pi-3\mu)\frac{k}{2}\right)\over
2\cosh{\mu k\over{2}}\sinh\left((\pi-2\mu)\frac{k}{2}\right)} \,, 
\label{Gspin1} \\
\widehat G_2(k)&=&{e^{-{\mu k\over{2}}}\over{e^{{\mu 
k\over{2}}}+e^{-{\mu k\over{2}}}}}\,,
\label{G2def}\\
C(k) &=& -\widehat G(k)(\widehat{Lt''_+}(k)+\widehat{{\cal L}t''_-}(k))
+ \widehat G_2(k)\, \widehat{Lf''}(k) +\widehat{LC''_b}(k)
- e^{-{\mu k\over 2}}\widehat{L\kappa''}(k) \non \\
&+& {2\pi k\over e^{\mu k\over 2}+e^{-{\mu k\over 2}}} \non\\
&=& C_{N}(k) + C_{1}(k) + C_{\tilde{B}_{\pm}}(k)
\label{spin1C}
\eeqa
where the expressions for $C_{N}(k)$, $C_{1}(k)$, and $C_{\tilde{B}_{\pm}}(k)$ (evaluated using (\ref{f}),(\ref{Cb}) and (\ref{tpm})) are given below,
\beqa
C_{N}(k) &=& 2\pi N \psi(k)(e^{i\La k} + e^{-i\La k})\big[-\widehat{G}(k)(e^{-\frac{\mu k}{2}} + e^{-\frac{3\mu k}{2}} - e^{-k\pi +\frac{\mu k}{2}} - e^{-k\pi +\frac{3\mu k}{2}})\non \\ 
&+& \widehat{G}_{2}(k)(e^{-\frac{3\mu k}{2}} + e^{-k\pi +\frac{3\mu k}{2}}) + (e^{-\frac{\mu k}{2}} - e^{-k\pi +\frac{\mu k}{2}} - e^{-k\pi +\frac{3\mu k}{2}})\big]\,, \label{CN}\\
C_{1}(k) &=& 2\pi\psi_{2}(k)\big[-\widehat{G}(k)(e^{-k\mu} - e^{k\mu - \frac{k\pi}{2}})+ \widehat{G}_{2}(k)(e^{-k\mu} + e^{k\mu - \frac{k\pi}{2}})\big] + 2\pi(k\widehat{G}(k) - \psi_{2}(k)e^{k\mu-\frac{k\pi}{2}}\non\\
&+& \frac{k}{e^{\frac{k\mu}{2}} + e^{-\frac{k\mu}{2}}})\,, \label{C1}\\
C_{\tilde{B}_{\pm}}(k) &=& -2\pi\psi(k)\Big\{
s_{-}e^{-\left(-{\mu |A_{-}|\over{2}}+\pi\right)k}+s_{+}e^{-\left(-{\mu |A_{+}|\over{2}}+\pi\right)k}
-s_{-}e^{-{\mu |A_{-}|\over{2}}k}-s_{+}e^{-{\mu |A_{+}|\over{2}}k} \non \\
&+&
e^{-\left(\mu B_{-}+\pi\right){k\over 2}}+e^{-\left(\mu B_{+}+\pi\right){k\over 2}}-e^{-\left(-\mu B_{-}+\pi\right){k\over 2}}-e^{-\left(-\mu B_{+}+\pi\right){k\over 2}}\Big\}\non\\
&\times & \big( \frac{1+e^{k\mu}+e^{-k\mu}}{e^{\frac{k\mu}{2}} + e^{-\frac{k\mu}{2}}} -\widehat{G}(k)(e^{\frac{k\mu}{2}} + e^{-\frac{k\mu}{2}})\big) 
\label{CBpm}
\eeqa
where $s_{\pm}\equiv$ sgn($A_{\pm}$), $\psi(k) \equiv {k\over 1-e^{-\pi k}}$ and  $\psi_2(k) \equiv {k\over 1-e^{-{\pi k\over 2}}}$
and we have used the following identities (see also \cite{ANS}), 
\beq
\int_{C_2} {du\over 2\pi}  \left[ \ln \sinh(u-i \alpha)\right]'' e^{iku}= 
e^{-k(\alpha - n \pi)}  \psi(k) \,, 
\label{psi}
\eeq
where $n$ is an integer such that $0 < \Re e(\alpha - n \pi) < \pi$, 
and
\beq
\int_{C_2} {du\over 2\pi} \left[ \ln \sinh(2u)\right]'' e^{iku}
=\psi_2(k) \,.
\label{psi2}
\eeq
Evaluation of $C(k)$ from (\ref{spin1C}) thus yields the following,
\beqa
C(k) &=& 2\pi k \Bigg\{
N\left({e^{i\La k}+e^{-i\La k}\over{2\cosh{\mu k\over{2}}}}\right)
+ {s_{+}\sinh\left((\pi-\mu |A_{+}|)\frac{k}{2}\right)
   +s_{-}\sinh\left((\pi-\mu |A_{-}|)\frac{k}{2}\right)
\over 2\cosh({\mu k\over 2}) \sinh\left((\pi-2\mu)\frac{k}{2}\right)} 
\non \\
& & + {\sinh\left(\frac{k}{2}\mu B_{+}\right)
   +\sinh\left(\frac{k}{2}\mu B_{-}\right)
\over 2\cosh({\mu k\over 2}) \sinh\left((\pi-2\mu)\frac{k}{2}\right)}
+{\cosh \frac{\mu k}{4} \sinh\left((3\mu -\pi)\frac{k}{4}\right)\over
\cosh \frac{\mu k}{2}\sinh\left((2\mu -\pi)\frac{k}{4}\right)}
\Bigg\} \,.
\label{Ck}
\eeqa

\noindent The third NLIE involves $y(u)$. 
\beq
\widehat{L y''}(k)=\widehat{L\check{\Lambda}''_2}(k) + \int_{C_2} du\ [\ln \sinh(2u)]'' e^{iku} 
+ \widehat{L\kappa''}(k) -\widehat{Lf''}(k) \,.
\label{ftLy}
\eeq
To this end, we proceed with the evaluation of $\widehat{L\check{\Lambda}''_2}(k)$. From (\ref{cauchy1}) and (\ref{cauchy1b}), and through the elimination of 
$\widehat{Lq''}(k)$, we have 
\beq
\left(e^{{\mu k\over{2}}}+e^{-{\mu k\over{2}}}\right)
\widehat{L\check{\Lambda}''_2}(k)=e^{-{\mu k\over{2}}}\widehat{L{\bar B}''}(k)-
e^{{\mu k\over{2}}}\widehat{{\cal L}B''}(k)+e^{-{\mu k\over{2}}}\widehat{Lt''_+}(k) -
e^{{\mu k\over{2}}}\widehat{{\cal L}t''_-}(k) \,.
\label{ftT2result}
\eeq
Solving for $\widehat{L\check{\Lambda}''_2}(k)$ from (\ref{ftT2result}), and using the result in (\ref{ftLy}), we obtain
\be
\widehat{Ly''}(k) = -\widehat G(-k)\, \widehat{{\cal L}B''}(k)+
\widehat G_2(k)\, \widehat{L{\bar B}''}(k)+ C_y(k) \,,
\label{nlie1}
\ee
where
\be
C_y(k)={e^{-{\mu k\over{2}}}\widehat{Lt''_+}(k) -
e^{{\mu k\over{2}}}\widehat{{\cal L}t''_-}(k)\over{e^{{\mu k\over{2}}}+e^{-{\mu k\over{2}}}}}
+ 2\pi \psi_{2}(k) + \widehat{L\kappa''}(k)-\widehat{Lf''}(k) \,,
\label{Cydef}
\ee
which after some manipulation yields the following,
\be
C_y(k)= 4\pi k \widehat G_2(-k) \,. \label{Cyresult}
\ee
Thus, (\ref{Lbnlie}) and (\ref{nlie1}) represent the NLIEs of the inhomogeneous lattice model of the spin-$1$ XXZ 
chain with general integrable boundary terms in Fourier space. 
For further computation in subsequent sections, we express these NLIEs in coordinate space. By integrating twice, we obtain 
\be
\ln b(u) &=& 
\int_{-\infty}^{\infty}du'\ G(u-u'-i\epsilon) 
\ln (1 + b(u'+i\epsilon)) 
-\int_{-\infty}^{\infty}du'\ G(u-u'+i\epsilon) \ln (1+ \bar 
b(u'-i\epsilon))\non \\
&+& \int_{-\infty}^{\infty}du'\ G_{2}(u-u'+i\epsilon) \ln 
(1+ y(u'-i\epsilon)) + i 2N \tan^{-1}\left({\sinh \frac{\pi u}{\mu}\over 
\cosh \frac {\pi \La}{\mu} } \right) \non \\
&+& i\, \int_{0}^{u}du'\ R(u')  + Ci\pi\,, 
\non \\
\ln \bar b(u) &=& 
-\int_{-\infty}^{\infty}du'\ G(u-u'-i\epsilon) 
\ln (1 + b(u'+i\epsilon)) 
+\int_{-\infty}^{\infty}du'\ G(u-u'+i\epsilon) \ln (1 + \bar 
b(u'-i\epsilon))\non \\
&+& \int_{-\infty}^{\infty}du'\ G_{2}(u'-u+i\epsilon) \ln 
(1 + y(u'+i\epsilon)) - i 2N \tan^{-1}\left({\sinh \frac{\pi u}{\mu}\over 
\cosh \frac {\pi \La}{\mu} } \right) \non \\
&-& i\, \int_{0}^{u}du'\ R(u')  - Ci\pi\,, \non \\
\ln y(u) &=& \int_{-\infty}^{\infty}du'\ 
G_{2}(u-u'+i\epsilon) \ln (1 + \bar 
 b(u'-i\epsilon)) + \int_{-\infty}^{\infty}du'\ 
G_{2}(u'-u+i\epsilon) \ln (1 + 
b(u'+i\epsilon))  \non \\
&+& 4\pi i\, \int_{0}^{u}du'\ G_{2}(-u') + C_{y}i\pi \,. 
\label{NLIEspin1u}
\ee
where $R(u)$ refers to the Fourier transform of $\hat R(k)$ which is given below,
\be
\hat R(k) &=& 2\pi \Bigg\{  
{s_{+}\sinh\left((\pi-\mu |A_{+}|)\frac{k}{2}\right)
   +s_{-}\sinh\left((\pi-\mu |A_{-}|)\frac{k}{2}\right)
\over 2\cosh({\mu k\over 2}) \sinh\left((\pi-2\mu)\frac{k}{2}\right)} 
\non \\
& & + {\sinh\left(\frac{k}{2}\mu B_{+}\right)
   +\sinh\left(\frac{k}{2}\mu B_{-}\right)
\over 2\cosh({\mu k\over 2}) \sinh\left((\pi-2\mu)\frac{k}{2}\right)} + {\cosh({\mu k\over 4}) \sinh\left((3\mu-\pi)\frac{k}{4}\right)\over
  \cosh({\mu k\over 2}) \sinh\left((2\mu-\pi)\frac{k}{4}\right)} 
\Bigg\} \,.
\label{Rk}
\ee
The terms $\pm Ci\pi$ and $C_{y}i\pi$ in (\ref{NLIEspin1u}) are integration constants. These factors are obtained by considering the $u\rightarrow\infty$ limit of 
(\ref{bb}), (\ref{y}) and (\ref{NLIEspin1u}). 
Proceeding as in \cite{ANS}, one obtains $C = -1$ and $C_{y} = 0$. We have explicitly checked that this procedure yields the same integration constant for all 
possible combinations of  the boundary parameter values, namely all four regions given in (\ref{regions}).

\section{Boundary SSG model and the continuum limit }\label{sec:SSG}

To make the paper relatively self-contained, we first give a brief review of the boundary SSG model (mainly reproduced from \cite{NepSSG}). This is followed by the treatment of the continuum 
limit of the NLIEs given in (\ref{NLIEspin1u}). 

\subsection{The boundary SSG model}\label{sec: bSSG}

The Euclidean-space action of the boundary SSG model is given by
\be
S &=& \int_{-\infty}^{\infty} dy \int_{-\infty}^{0} dx\ {\cal L}_{0}
+ \int_{-\infty}^{\infty} dy\  {\cal L}_{b} \,,
\ee
where the bulk Lagrangian density is given by
\be
{\cal L}_{0} =
2 \partial_{z}\varphi \partial_{\bar z} \varphi  
- 2 \bar \psi \partial_{z} \bar \psi 
+ 2 \psi \partial_{\bar z} \psi 
- 4 \cos \varphi - 4 \bar \psi \psi \cos {\varphi\over 2} \,,
\label{bulkL}
\ee
In (\ref{bulkL}), $\psi$ and $\bar \psi$ are the two components of a Majorana
Fermion field, and $z=x+iy$, $\bar z=x-iy$.  The boundary Lagrangian at $x=0$ is given by
\be
{\cal L}_{b} = \bar \psi \psi + ia \partial_{y} a 
- 2 p(\varphi) a (\psi - \bar \psi) + {\cal B}(\varphi) \,,
\label{boundL}
\ee
where $a$ is a Hermitian Fermionic boundary degree of freedom. The functions $p(\varphi)$ and ${\cal B}(\varphi)$, which are
potentials that are functions of the scalar field $\varphi$, are determined from the requirement of boundary integrability and supersymmetry. They are found to be 
\cite{NepSSG}
\be
{\cal B}(\varphi) = 2 \upsilon \cos {1\over 2}(\varphi - \varphi_{0}) \,, \qquad p(\varphi) = {\sqrt{F}\over 2}\sin{1\over 4}(\varphi - D)\qquad 
\mbox{where} \quad 
\tan {D\over 2} ={\upsilon \sin{\varphi_{0}\over 2}\over
\upsilon \cos{\varphi_{0}\over 2} -4} \,.
\label{Bpotential}
\ee
where $F = \sqrt{\upsilon^{2}-8\upsilon\cos \frac{\varphi_{0}}{2}+16}$

The parameters $\upsilon$ and $\varphi_{0}$ are arbitrary and real.
As mentioned in \cite{ANS}, in the limit that the boundary mass parameters tend to infinity, one arrives at the SSG model with Dirichlet boundary conditions. This 
model corresponds to the open spin-$1$ XXZ chain with diagonal boundary terms \cite{MNRitt}. This case was considered in detail in \cite{ANS}.

\subsection{The continuum limit}\label{continuum}

Next, we proceed by analogy with the boundary SG case with general integrable boundary conditions \cite{AN} and the boundary SSG case with 
Dirichlet boundary conditions \cite{ANS}. 
There, in the continuum limit, which consists of taking $\La
\rightarrow \infty$, $N \rightarrow \infty$ and lattice spacing $\Delta
\rightarrow 0$, such that the interval length $L$
and the soliton mass $m$ are given by 
\be
L = N \Delta \,, \qquad m={2\over \Delta} e^{-{\pi \Lambda\over \mu}} \,,
\label{continuumlimit}
\ee
respectively, the NLIEs of the inhomogeneous open XXZ spin-$1/2$ and spin-$1$ chains are shown to describe the 
NLIEs of the boundary SG and SSG models respectively. It is therefore natural to conjecture that a set of NLIEs describing a boundary SSG model that corresponds 
to NLIEs of the inhomogeneous open XXZ spin-$1$ chain with general integrable boundary terms can also be derived here. In the continuum limit, the term 
$-i2N \tan^{-1}\left({\sinh \frac{\pi u}{\mu}\over 
\cosh \frac {\pi \La}{\mu}} \right)$ becomes $-i 2mL \sinh \theta$ after defining the renormalized rapidity $\theta$ as
\be 
\theta = \frac{\pi u}{\mu} \,.
\label{renormrapidity}
\ee 
Thus, (\ref{NLIEspin1u}) becomes
\be
\ln \bff(\theta) &=& 
\int_{-\infty}^{\infty}d\theta'\ \Gf(\theta-\theta'-i\varepsilon) 
\ln (1+\bff(\theta'+i\varepsilon)) 
-\int_{-\infty}^{\infty}d\theta'\ \Gf(\theta-\theta'+i\varepsilon) \ln (1+\bar 
\bff(\theta'-i\varepsilon))\non \\
&+& \int_{-\infty}^{\infty}d\theta'\ \Gf_{2}(\theta-\theta'+i\varepsilon) \ln 
(1+ \yf(\theta'-i\varepsilon)) + i 2mL \sinh \theta + i\,  
\Pf_{bdry}(\theta) -i\pi\,, 
\non \\
\ln \bar \bff(\theta) &=& 
-\int_{-\infty}^{\infty}d\theta'\ \Gf(\theta-\theta'-i\varepsilon) 
\ln (1+\bff(\theta'+i\varepsilon)) 
+\int_{-\infty}^{\infty}d\theta'\ \Gf(\theta-\theta'+i\varepsilon) \ln (1+\bar 
\bff(\theta'-i\varepsilon))\non \\
&+& \int_{-\infty}^{\infty}d\theta'\ \Gf_{2}(\theta'-\theta+i\varepsilon) \ln 
(1+\yf(\theta'+i\varepsilon)) - i 2mL \sinh \theta - i\,  
\Pf_{bdry}(\theta) + i\pi \,, \non \\
\ln \yf(\theta) &=& \int_{-\infty}^{\infty}d\theta'\ 
\Gf_{2}(\theta-\theta'+i\varepsilon) \ln 
(1+\bar \bff(\theta'-i\varepsilon)) + \int_{-\infty}^{\infty}d\theta'\ 
\Gf_{2}(\theta'-\theta+i\varepsilon) \ln 
(1+\bff(\theta'+i\varepsilon))  \non \\
&+& i\, \Pf_{y}(\theta) \,. 
\label{NLIEspin1coord}
\ee
where following definitions have been used,
\be
\varepsilon= \frac{\pi \epsilon}{\mu}\,, \quad \!
\bff(\theta)= b(\frac{\mu \theta}{\pi})\,, \quad \!
\yf(\theta)= y(\frac{\mu \theta}{\pi})
\label{mathfrakdefs}
\ee 
Moreover, $\Gf(\theta)$ and $\Gf_{2}(\theta)$ are given by
\be
\Gf(\theta) &=&  \frac{\mu}{\pi} G(\frac{\mu \theta}{\pi}) \non\\
&=& {\mu\over 2\pi^{2}} \int_{-\infty}^{\infty}dk\ e^{-i 
k \mu \theta/\pi}\ \widehat G(k) \,,
\ee
\be
\Gf_{2}(\theta) &=& \frac{\mu}{\pi} G_{2}(\frac{\mu \theta}{\pi}) \non\\
&=&  {\mu\over 2\pi^{2}} \int_{-\infty}^{\infty}dk\ e^{-i 
k \mu \theta/\pi}\ \widehat G_{2}(k) 
= {i\over 2\pi \sinh \theta} \,,
\label{G2theta}
\ee
where 
$\widehat G(k)$ and $\widehat G_{2}(k)$ are as given in (\ref{Gspin1}) and (\ref{G2def}) respectively.
From (\ref{NLIEspin1u}), we also note that $\Pf_{bdry}(\theta)$ and $\Pf_{y}(\theta)$  are given by
\be
\Pf_{bdry}(\theta) &=& P_{bdry}(\frac{\mu \theta}{\pi})\non \\
&=&  {\mu\over 4\pi^{2}} \int_{-\theta}^{\theta} d\theta' \int_{-\infty}^{\infty}dk\ 
e^{-i k \mu \theta'/\pi}\ \hat R(k)\,,
\label{pbdry}
\ee
and
\be
\Pf_{y}(\theta) &=& P_{y}(\frac{\mu \theta}{\pi})\non \\
&=& 4\pi \int_{-\infty}^{\theta}d\theta'\ 
\Gf_{2}(-\theta') = -2i \ln 
\tanh \frac{\theta}{2} - 2\pi \,,
\label{Pytheta}
\ee
respectively. Note that in (\ref{pbdry}), $\hat R(k)$ is given by (\ref{Rk}). The boundary terms $\Pf_{bdry}(\theta)$ and $\Pf_{y}(\theta)$\footnote{The function 
$\Pf_{y}(\theta)$ is the same as for the Dirichlet boundary conditions case found in \cite{ANS}.}
are essential in the investigations of IR limit (when computing the boundary $S$ matrix) and  
the UV limit (for the computation of effective central charge which is treated in section 6). Thus, in particular, $\Pf_{bdry}(\theta)$ is among the important results of 
this paper.  

\section{Boundary and Casimir energies}\label{sec:boundaryCasimir}

In this section, we compute the boundary correction (order 1) and the Casimir correction (order $1/L$) to the energy. Following the prescription given in \cite{RS}
(see also \cite{ANS}), according to which the energy for the inhomogeneous case ($\Lambda \ne 0$) is given by 
\beq
E=-\frac{g}{\Delta}\left\{ \frac{d}{du}\ln \tilde{\Lambda}_{2}(u)\Bigg\vert_{u=\Lambda+\frac{i\mu}{2}}
-\frac{d}{du}\ln \tilde{\Lambda}_{2}(u)\Bigg\vert_{u=\Lambda-\frac{i\mu}{2}}\right\} 
\,, \label{energydef}
\eeq
where $g$ is given by 
\be
g=-\frac{i\mu}{4\pi} \,. \label{energynormalization}
\ee
Using (\ref{y}) and the following,
\be
[\ln h(u)]' = \int \frac{dk}{2\pi}\ \widehat{Lh'}(k)\ e^{-ik u} \,, \qquad
u \in C_{2} \,,
\label{FTfact}
\ee 
(\ref{energydef}) reduces to
\be
E &=&-\frac{g}{\Delta}\int \frac{dk}{2\pi} e^{-ik\Lambda}
\big(e^{\frac{\mu k}{2}} - e^{-\frac{\mu k}{2}}\big)\left[\widehat{Ly'}(k)+\widehat{Lf '}(k)+\frac{2\pi \psi_{2}(k)}{ik}\right] \,,
\label{FTenergy}
\ee
where $\psi_{2}(k)$ is as defined in section 3. We have also used the following result,
\be
\widehat{Lh'}(k) = {1\over (-i k)}\widehat{Lh''}(k)
\label{fprime}
\ee
By evaluating the Fourier transform of (\ref{f}) and using (\ref{nlie1}) together with (\ref{fprime}), we can write the energy $E$ as,
\be
E = E_{L} + E_{1} + E_{1/L}
\label{energybulkboubdCas}
\ee
where the $E_{L}$, $E_{1}$ and $E_{1/L}$ represent the bulk vacuum energy, boundary vacuum energy and Casimir energy respectively:
\beq
E_{L}=\frac{2Ng}{i \Delta}\int_{-\infty}^{\infty} dk e^{-2i\Lambda k}
{\sinh \frac{\mu k}{2}\cosh\left((\frac{3\mu}{2}-\frac{\pi}{2})k\right)
\over \sinh\frac{\pi k}{2}} \,.
\label{bulk}
\eeq
\be
E_{1}&=&\frac{2g}{i \Delta}\int_{-\infty}^{\infty} dk e^{-i\Lambda k}
\sinh\left({\mu k\over{2}}\right)\Bigg\{\frac{e^\frac{\mu k}{2}}{\ch \frac{\mu k}{2}}
+ \frac{{\cosh\left(\frac{\pi k}{2}-s_{+}\frac{\mu k}{2}(A_{+}+1)\right)}}{\sinh \frac{\pi k}{2}} \non \\
&+& \frac{{\cosh\left((B_{+}-1)\frac{k\mu}{2} \right)}}{\sinh \frac{\pi k}{2}} 
+ (+ \leftrightarrow -)
+ \frac{e^{(\frac{\pi}{4}-\mu )k}
+e^{(\mu - \frac{\pi}{4})k}-e^{\frac{\pi k}{4}}}{2\sinh{\pi k\over 4}}
\Bigg\}\,,
\label{boundary}
\ee
\beq
E_{1/L}=-\frac{g}{2\Delta}\int_{-\infty}^{\infty} \frac{dk}{2\pi} e^{-i\Lambda k}
\frac{1}{\cosh{\mu k\over{2}}}\left[
\widehat{L{\bar B}'}(k)+\widehat{{\cal L}B'}(k) - e^{-k\mu}\widehat{L{\bar B}'}(k) - e^{k\mu}\widehat{{\cal L}B'}(k)\right] \,.
\label{Casimir}
\eeq
In (\ref{boundary}), the symbol $(+\leftrightarrow -)$ represents the terms with 
$A_{+}\rightarrow A_{-}\,, B_{+}\rightarrow B_{-}\,, s_{+}\rightarrow s_{-}$.
In (\ref{bulk}), adopting the renormalization procedure \cite{DdV}-\cite{DdV3} of
keeping only the finite terms and by closing the integral contour in the lower half plane, selecting only the contribution 
from the residue at $k = -i\frac{\pi}{\mu}$, we note that the integrand is analytic on and within the contour in the lower plane, hence
giving
\be
E_{L}=0 \,, \label{bulkvacuumenergy}
\ee
This is in agreement with known results (see, e.g., \cite{Du2}). Also see \cite{ANS}. As found in \cite {ANS}, for the case with Dirichlet boundary conditions, next, 
by using the same contour,  we find that the boundary vacuum energy (\ref{boundary}), is given by
\be
E_{1}=m \,. \label{SSGboundvacuumenergy}
\ee
where $m$ is given by (\ref{continuumlimit}). Hence each boundary contributes the vacuum energy $m/2$, and it is independent of the boundary parameters.
Finally, the Casimir energy (\ref{Casimir}) simplifies to
\beq
E_{1/L}=\frac{m}{2\pi}\Im m\int_{-\infty}^{\infty} d\th\
\sinh(\th-i\varepsilon)\ln (1+\bar \bff(\th-i\varepsilon)) \,. 
\label{Casimirspin1}
\eeq

\section{UV limit}\label{sec:UV}

We now consider the Casimir energy in the UV limit $m L \rightarrow 0$. A systematic approach utilizing dilogarithm trick \cite{Su1, KP, KP2, KP3} 
was used in \cite{ANS} for the SSG models with Dirichlet boundary conditions. The analysis of our set of equations is similar to the one presented there. Thus,
we give only the significant differences with respect to \cite{ANS}. Denoting the reference \cite{ANS} here by I, the $\Pf_{bdry}(\infty)$, (I5.5) becomes,  
\be
\Pf_{bdry}(\infty) &=& \frac{1}{2} \hat R(0)\non\\
&=& \pi\bigg(1 + \frac{\pi(s_{+} + s_{-})}{4(\frac{\pi}{2}-\mu)} -  \frac{\mu(A_{+} + A_{-})}{4(\frac{\pi}{2}-\mu)}\bigg)
 \,.
\label{Pifnty}
\ee 
The quantity (I5.15) becomes,
\be
\omega = \frac{\Pf_{bdry}(\infty) - \pi}{2\left[\frac{3}{4}-G(\infty) 
\right]} = \pi\left(s_{+}+s_{-}\right) - \mu\left( A_{+}+A_{-}\right)  \,, \label{omega}
\ee
where $G(\infty) = \frac{\pi-3\mu}{2(\pi-2\mu)}$. In (\ref{omega}), $\omega$ is required to obey the bound $|\omega| < \frac{2\pi}{3}$.
We further remark that the regions of the boundary parameters $A_{\pm}$ (\ref{regions}), 
should be restricted further as follows in order for this bound to be satisfied:
\be
\begin{array}{r@{\ : \quad}l}
    I   & \frac{4\pi}{3\mu} < (A_{+} + A_{-}) < \frac{8\pi}{3\mu} \\
    II, IV  & -\frac{2\pi}{3\mu} < (A_{+} + A_{-}) < \frac{2\pi}{3\mu} \\
    III &  -\frac{8\pi}{3\mu} < (A_{+} + A_{-}) < -\frac{4\pi}{3\mu} \\
      \end{array}
  \label{regions2}
\ee
The Casimir energy (I5.17) is thus given by
\be
L\, E_{1/L}(0) = -\frac{\pi}{24} \left[ \frac{3}{2} - 
\frac{3}{\pi (\pi-2\mu)} 
\left(\pi(s_{+}+s_{-})-\mu(A_{+}+A_{-})\right)^{2} \right] \,.
\label{Casimirresult}
\ee 
The above result can be rewritten in terms of the effective central charge $c_{eff}(0)$,
\be
E_{1/L}(0)=-\frac{\pi}{24L}c_{eff}(0)   \,,
\label{ceffective}
\ee 
Hence, the result (\ref{Casimirresult}) implies that the effective central 
charge (I5.18) becomes
\be
c_{eff}(0) = \frac{3}{2} - 
\frac{3}{\pi (\pi-2\mu)} 
\left[\pi(s_{+}+s_{-})-\mu (A_{+}+A_{-})\right]^{2} \,,
\label{ceffresult}
\ee
which after imposing constraints (\ref{realconstraints}), with $k=1$, becomes
\be
c_{eff}(0) = \frac{3}{2} - 
\frac{3}{\pi (\pi-2\mu)} 
\left[\pi(s_{+}+s_{-})\mp\mu |c_{-}-c_{+}|\right]^{2} \,,
\label{ceffresult2}
\ee

We note that the terms in square bracket in (\ref{ceffresult}) and (\ref{ceffresult2}) also appear in the SG case with general integrable boundary conditions
(refer to equations (2.33) and (2.34) in \cite{AN}). We also note the similarity with the results obtained in \cite{ANS}, where the boundary SSG model with
Dirichlet boundary conditions reduces to a system of one free Boson and one free Majorana fermion, the central charge of which is given by \cite{Saleurlecture},
$c = c_{B} + c_{F} = 1 + \frac{1}{2} = \frac{3}{2}$.

\section {Discussion}\label{sec:dis}

Starting from the $T$-$Q$ equations for the spin-$1$ XXZ chain, we derived the NLIEs for inhomogeneous spin-$1$
XXZ chain with general integrable boundary terms, satisfying a certain constraint (\ref{constraint}). We have restricted our
analysis here to the ground state. Further, taking the continuum limit, and working in analogy with the usual SG model and the SSG
model on an interval (with Dirichlet boundary conditions), we also obtain the NLIEs that should describe the corresponding
SSG model. We further compute the energies that include bulk, boundary and Casimir terms. As expected, the bulk contribution vanishes. The
boundary energy turns out to be independent of the lattice boundary parameters, ($A_{\pm}\,, B_{\pm}$) (as for the case with Dirichlet boundary conditions). We also 
looked at the UV limit of the Casimir energy and obtained an expression for the effective central charge (\ref{ceffresult}), that has the right $c = 3/2$ factor as expected. 

One could also further investigate this model. For example, the IR limit can be studied. The result obtained for $\Pf_{bdry}(\theta)$ is useful for such
an analysis. In addition, one could also investigate the excited states of the model. Last but not least, it should also be possible to carry out such analysis
for open spin-$1$ XXZ chain with non-diagonal boundary terms, the solutions to which are given in \cite{M2}. In contrast to the solution used in this paper, 
the solutions given in \cite{M2} are not restricted by any type of constraints among the lattice boundary parameters. It would be interesting to work out the 
NLIEs for these cases as well and analyze their UV and IR limits. We hope to address some of these issues in future.

\end{document}